\documentclass[journal,twoside,web]{ieeecolor}
\usepackage{jsen}
\usepackage{cite}
\usepackage{amsmath,amssymb,amsfonts}
\usepackage{algorithmic}
\usepackage{graphicx}
\usepackage{textcomp}
\usepackage{wrapfig}
\usepackage{multirow}
\usepackage{hyperref}

\def\BibTeX{{\rm B\kern-.05em{\sc i\kern-.025em b}\kern-.08em
    T\kern-.1667em\lower.7ex\hbox{E}\kern-.125emX}}
\definecolor{abstractbg}{rgb}{0.89804,0.94510,0.83137}
\begin{document}
\title{Cognitive Inference based Feature Pyramid Network for Sentimental Analysis using EEG Signals}
\author{Vishesh Bhardwaj, Aman Yadav, Srikireddy Dhanunjay Reddy, Tharun Kumar Reddy Bollu
\thanks{V. Bhardwaj, A. Yadav, S. D. Reddy, and T. K. Reddy Bollu are with 
the Electronics \& Communication Engineering Department, Indian Institute of Technology, Roorkee, India. (e-mail: \{vishesh\_b, aman\_y\}@ch.iitr.ac.in, \{sd\_reddy, tharun.reddy\}@ece.iitr.ac.in)}}
\IEEEtitleabstractindextext{%
\fcolorbox{abstractbg}{abstractbg}{%
\begin{minipage}{\textwidth}%
\begin{wrapfigure}[7]{r}{4in}%
\includegraphics[width=3.9in]{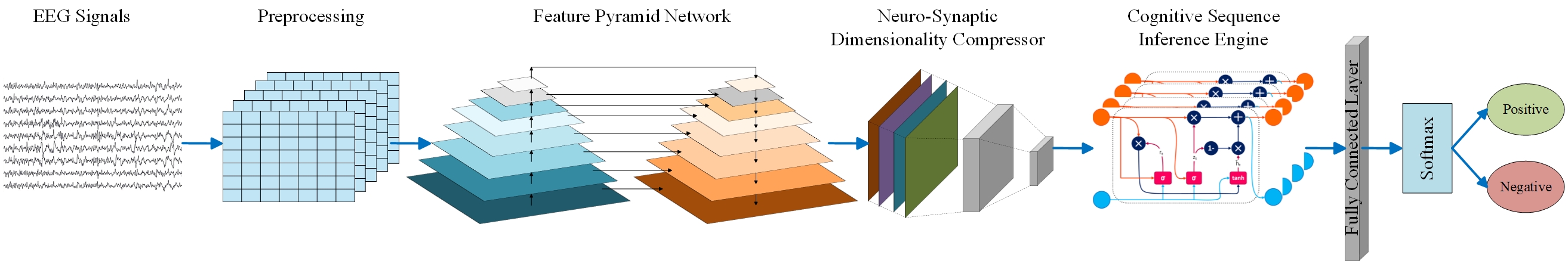}%
\end{wrapfigure}%
\begin{abstract}
Sentiment analysis using Electroencephalography (EEG) sensor signals provides a deeper behavioral understanding of a person's emotional state, offering insights into real-time mood fluctuations. This approach takes advantage of brain electrical activity, making it a promising tool for various applications, including mental health monitoring, affective computing, and personalised user experiences. An encoder-based model for EEG-to-sentiment analysis, utilizing the ZUCO 2.0 dataset and incorporating a Feature Pyramid Network (FPN), is proposed to enhance this process. FPNs are adapted here for EEG sensor data, enabling multiscale feature extraction to capture local and global sentiment-related patterns. The raw EEG sensor data from the ZUCO 2.0 dataset is pre-processed and passed through the FPN, which extracts hierarchical features. In addition, extracted features are passed to a Gated Recurrent Unit (GRU) to model temporal dependencies, thereby enhancing the accuracy of sentiment classification. The ZUCO 2.0 dataset is utilized for its clear and detailed representation in 128 channels, offering rich spatial and temporal resolution. The experimental metric results show that the proposed architecture achieves a 6.88\% performance gain compared to the existing methods. Furthermore, the proposed framework demonstrated its efficacy on the validation datasets DEAP and SEED.
\end{abstract}
\end{minipage}}}
\maketitle 
\section{Introduction}
\label{sec:introduction}
EEG is a widely used non-invasive technique for monitoring brain activity, capturing the brain's electrical signals in real-time. These signals reflect various cognitive processes and emotional states, making EEG signal analysis a powerful tool in neuroscience research, clinical diagnostics, and even consumer applications such as brain-computer interfaces \cite{b1}. Traditionally, EEG signal analysis has relied on conventional signal processing techniques such as Fourier transforms, wavelet transforms, and power spectral density analysis to extract meaningful patterns \cite{b2}. In addition, some advanced techniques are implemented to understand the underlying connectivity among channels using topological approaches \cite{d1}, \cite{d2}, \cite{d3}. However, these methods often require handcrafted feature extraction, which may fail to capture the full complexity and temporal dependencies within the data. With the growing availability of large EEG datasets and advances in deep learning, the field has shifted towards automated, data-driven approaches that can uncover latent patterns in EEG signals without extensive feature engineering \cite{b3}. One such example of a dataset is ZUCO 2.0, which has been used in EEG-based language and sentiment understanding. It offers structured annotations with high spatial and temporal resolution, making it suitable for multiscale feature extraction \cite{b9}.\\
EEG data often consist of various types of noise from muscle movements, eye blinks, and environmental interference; therefore, the preprocessing step becomes crucial for further analysis. Previous studies have used various techniques for the removal of noise to enhance the quality of the EEG signal. Common methods include band-pass filtering, Independent Component Analysis(ICA), and wavelet decomposition. In addition, normalization techniques such as min-max scaling and z-score standardization have been used to ensure consistency between different EEG channels and subjects \cite{b4}.\\
Sentiment analysis, the task of detecting and classifying emotional states from EEG data, plays a crucial role in understanding human emotions and has diverse applications in fields such as mental health monitoring, human-computer interaction, and affective computing. It involves interpreting complex patterns of brain activity to infer the underlying emotional states, which can range from happiness and sadness to stress or excitement. The process typically requires the extraction of relevant features from EEG signals, followed by classification into predefined emotional categories \cite{b5}. Traditional approaches have relied on methods such as convolutional neural networks (CNNs) and recurrent neural networks (RNNs), including long short-term memory (LSTM) networks, to capture spatial and temporal dependencies in the data \cite{b6}, \cite{b7}, \cite{b8}. However, these techniques have limitations. CNNs are effective in learning spatial features from EEG signals, but often do not fully capture the temporal dynamics of the data \cite{b6}. Meanwhile, RNN-based models like LSTMs, while adept at handling sequential dependencies, struggle with computational efficiency and issues such as vanishing gradients, which can hinder their performance on large datasets or long sequences \cite{b7}. \\
Given these challenges, there is a growing need for advanced architectures capable of extracting hierarchical multiscale features from EEG data while efficiently modeling temporal dependencies. FPN have shown great potential in this regard. Zheng et al. \cite{b10} proposed a collaborative and multilevel feature selection network that takes advantage of the FPN-based architecture to enhance action recognition by extracting discriminative characteristics at multiple spatial scales and levels. This approach demonstrates the effectiveness of FPNs in learning robust hierarchical characteristics, which can be adapted for EEG-based applications.
For temporal modeling, Ravanelli et al. \cite{b11} introduced Light Gated Recurrent Units (Li-GRUs), a streamlined variant of traditional GRUs optimized for speech recognition tasks. Their model significantly reduces computational overhead while preserving high accuracy, showcasing its potential for sequence modeling in real-time applications. These innovations underscore the adaptability of both FPN and GRU-based architectures in various domains and inspire their application in EEG-based sentiment analysis. Our framework introduces three key novel components that differentiate it from existing approaches:
\begin{itemize}
    \item Proposed a novel cognitive feature pyramid network framework, enabling multiscale feature extraction to capture both high-level abstractions and fine-grained emotional patterns in EEG signals.
    \item Implementation of FPN for the extraction of high and low-level features from EEG sensor data, facilitating superior performance through improved implementation of diverse feature maps of EEG data.
    \item Implementation of a Cognitive Sequence Inference Engine for EEG-based sentiment analysis, enabling efficient capture of temporal dynamics and effective handling of large-scale data streams to enhance overall predictive accuracy.
\end{itemize}
The rest of this article is organized as follows. Section II delves into the related studies that were done. Section III defines the data set used for sentiment analysis with its parameters. In addition, it explains the functional details of each and every block in the framework. Section IV explains the results obtained from the proposed model and compares them with existing models. Section V summarizes the paper with a brief discussion of the proposed framework and its performance.
\section{Related Studies}
Although many studies have been conducted using EEG datasets, only a few have focused on proper model-based modifications. Various modifications can be applied to the entire pipeline, as explained below.
\subsection{Feature Extraction Techniques for EEG Datasets}
Feature extraction for EEG datasets traditionally relies on frequency-domain analysis or time-domain features, such as power spectral densities. However, newer approaches leverage more advanced techniques such as wavelet transforms, sparse PCA, spatio-temporal field, and deep learning models to extract more useful characteristics \cite{b12}. A key challenge remains in selecting the most discriminative features that can capture the emotional states represented in the EEG data. Fuzzy logic models have brought about a revolutionary change in the accuracy of the models, with some achieving up to 98\% accuracy, compared to previous traditional models. This has recently emerged as a promising approach in signal processing. Unlike traditional models that rely on crisp values, fuzzy models allow for more flexible and nuanced decision-making, improving the classification of emotional states from EEG signals\cite{b13}.
\subsection{Emotion Recognition Using EEG Signals}
Bazgir et al. \cite{b14} analyzed EEG signals for emotion recognition using machine learning models. They used feature extraction techniques, such as wavelet transform, to decompose EEG signals into different frequency bands, and then machine learning classifiers, such as SVM and KNN, were used for multiclass classification. Similarly, Rahman et al. \cite{b15} explored a similar approach using Chi-square and sparsePCA for feature selection and extraction, followed by Light Gradient Boosting Machines (LGBM) for classification. Their work emphasized the importance of choosing the right features and classifiers to optimize EEG emotion recognition performance. Li et al. (2021) \cite{b16} propose a novel spatio-temporal field for emotion recognition based on EEG signals, which integrates both spatial and temporal features by using Rational Asymmetry of Spectral Power features to improve the accuracy of emotion classification. These methods highlight the importance of selecting the right features that capture emotional nuances in the EEG data, which is a key to improving the accuracy of the classification.
\subsection{Deep Learning Approaches for EEG-Based Emotion Analysis}
Deep learning techniques have significantly advanced emotion recognition from EEG signals by automating the feature extraction process and improving the model’s ability to learn complex patterns in the data. Cheng et al. \cite{b17} proposed the Attentional Temporal Convolutional Network (ATCN) to capture unimodal temporal features from EEG data. Their approach integrates temporal convolutional layers with attention mechanisms, allowing the network to focus on important time-dependent features and suppress irrelevant data. In contrast, the work on recognizing riding feelings from EEG introduced the use of recurring neural networks (RNNs) over spatio-temporal data, which allows for better modeling of temporal dependencies and spatial patterns in EEG signals\cite{b18}. Furthermore, the hierarchical spatial information learning model using Transformers for EEG-based emotion recognition takes advantage of the long-term memory capabilities of the model, allowing it to maintain and remember long-term dependencies across time and spatial dimensions for more accurate emotion classification \cite{b19}.
\subsection{Multimodal Sentiment Analysis and Fusion Models}
Although single-modality emotion recognition has made significant progress, the integration of multiple modalities such as EEG, facial expressions, and speech has shown superior performance in sentiment analysis. Soleymani et al. \cite{b20} explored multimodal sentiment analysis by combining EEG data with facial expressions. Their work demonstrated that the integration of multiple data sources enhances classification accuracy by leveraging complementary information. Similarly, Song et al. \cite{b21} proposed a graph-based method to capture the relationships between different EEG channels for emotion recognition. Although this approach has proven effective, many existing multimodal fusion models do not fully exploit the interactions between modalities. Cheng et al. \cite{b17} introduced a model that addresses this gap by combining attention mechanisms to capture both intramodal and intermodal interactions, ensuring that the unique characteristics of each modality are preserved while enhancing the overall performance of sentiment analysis.\\
These studies highlight the growing trend of integrating multiple data sources and sophisticated neural architectures for more robust and accurate emotion recognition systems. By combining techniques such as deep learning, fuzzy logic, and multimodal fusion, these advances lay the foundation for more accurate and real-time emotion detection systems.
\section{METHODS}
\subsection{Dataset Description}
\begin{figure*}[!t] 
    \centering
    \includegraphics[width=1\textwidth]{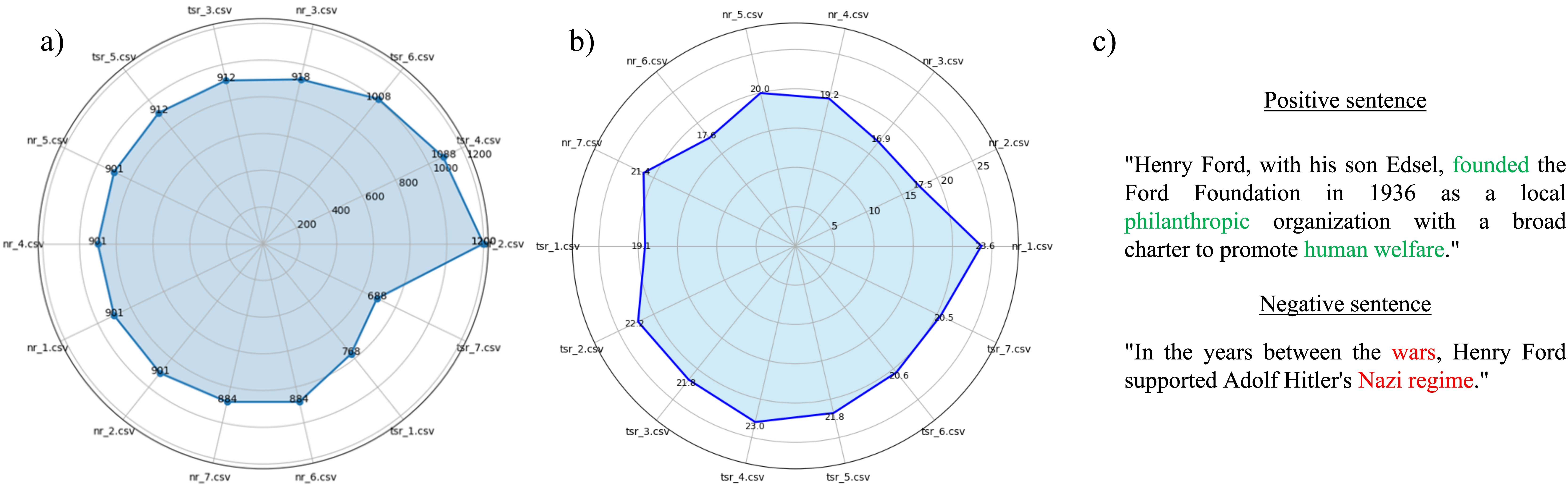} 
    \caption{ZUCO 2.0 dataset Feature a) Average Sentence Length (in words) b) Number of Words per file c) Example of Positive and Negative sentences }
    \label{fig 1:data description}
\end{figure*}

The EEG data in the ZUCO 2.0 dataset is a scientifically structured collection that provides an exceptional opportunity to study the neural underpinnings of reading and language comprehension from experimentation in a total of 18 subjects with a variety of backgrounds. It includes high-density recordings from 128 EEG channels, which capture brain activity with millisecond temporal resolution during various cognitive and linguistic tasks in the experimental setting. Figure \ref{fig 1:data description} represents a subject-wise comparison for the length of sentences and words and also represents an example of positive and negative sentences from the dataset. This multichannel setup ensures comprehensive spatial coverage of neural signals, allowing for an in-depth exploration of the brain regions involved in natural reading and other language processing tasks.\\
The ZUCO 2.0 dataset offers an extensive multichannel setup for EEG recordings, providing a detailed view of brain activity across various frequency bands and event-related potentials (ERPs). These frequency bands - Delta (0.5-4 Hz), Theta (4-8 Hz), Alpha (8-12 Hz), and Beta (12-30 Hz) - capture different aspects of cognitive and emotional states, with Theta and Alpha associated with memory and focus, and Beta associated with active mental engagement. The data set also tracks ERPs such as the P300, a key indicator of attention and emotional response. Each trial is carefully annotated with event markers, such as word onsets and sentence characteristics, allowing researchers to explore how the brain reacts to specific words or phrases. This comprehensive structure supports a broad range of cognitive and linguistic experiments, from analyzing word-level neural responses to examining sentence-level comprehension, making it a powerful tool for understanding language processing and cognitive functions. Figure \ref{fig 1:data description} shows the detailed description of the sentences given to the subjects during the recording of the EEG signal.

\subsection{Feature Pyramid Network}
\begin{figure}[!ht]
    \centering
    \includegraphics[width=1\columnwidth]{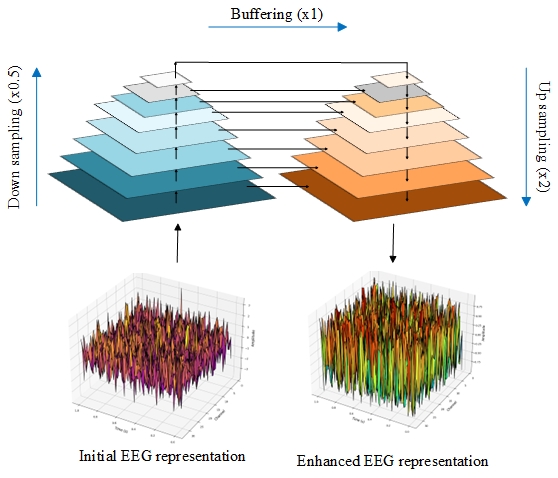}
    \caption{Schematic Diagram of FPN-based processing}
    \label{feature pyramid}
\end{figure}

FPNs are used to analyze EEG data from the ZUCO 2.0 dataset, addressing the complex and multiscale nature of brain signals. FPN offers a novel approach by enhancing the representation of EEG data, focusing on amplifying the most informative portions of the signals. Its hierarchical processing enables the model to effectively capture both fine-grained and high-level features, which are critical for understanding neural activity. The primary goal is to weigh the essential components of the EEG signals while retaining their spatial and temporal integrity.\\
The FPN processes the EEG data through a sequence of convolutions, starting with downsampling to reduce spatial dimensions and extract essential features. Each convolution layer reduces the feature map size by half using strided convolution with a stride equal to 2, mathematically expressed as:
\begin{equation}
    \mathbf{X}' = \frac{\mathbf{X}}{2}
\end{equation}
To enhance the representation, the network up-samples the feature map using padding and combines it with feature maps from earlier layers through element-wise addition. The enhanced feature representation is computed as
\begin{equation}
    F_{\text{Feature Map}} = F_{\text{Upsample}} + F_{\text{Downsample}}
\end{equation}
This mechanism ensures that both coarse and detailed features are preserved. Figure \ref{feature pyramid} shows the representation of the FPN structure with enhancement of the EEG signal characteristics.
The proposed model leverages advanced neural architectures to bridge EEG data and sentiment analysis tasks. By processing EEG signals through an FPN for feature extraction and a GRU module for sequence learning, the system achieves robust predictions. This method is in line with recent advances in multimodal sentiment analysis, which integrates the temporal and spatial dynamics of neural data. The modular design of the framework enhances interpretability and scalability, catering to diverse applications in cognitive neuroscience and emotion recognition.\\
The stride, padding, and number of layers are carefully adjusted according to the size of the data set to achieve optimal representation. The depth of upsampling and downsampling layers directly influences the network's ability to detect intricate features, effectively increasing the feature complexity. This refined representation serves as a highly reliable input for the subsequent stages of the proposed network, enabling a precise and nuanced analysis of the EEG signals.
\subsection{Proposed Framework}
\begin{figure*}[!t] 
    \centering
    \includegraphics[width=\textwidth]{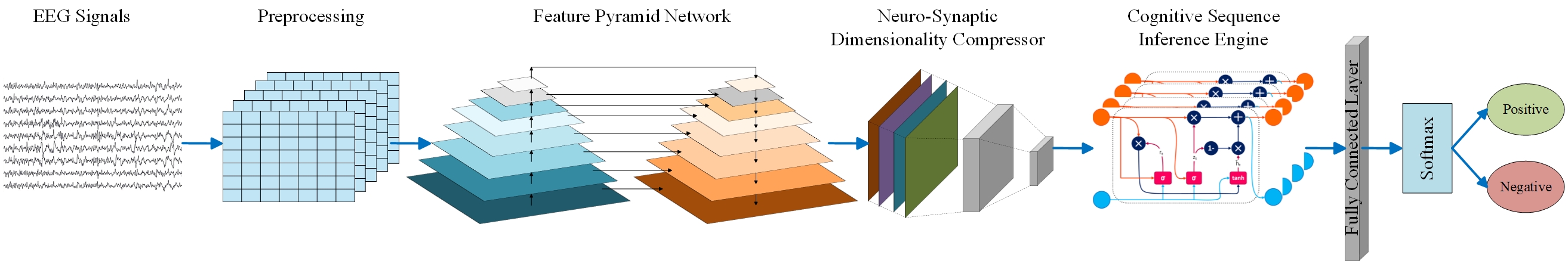}
    \caption{ Proposed framework for the EEG to sentiment prediction using cognitive inference based  feature pyramid network}
    \label{fig 3:methodology}
\end{figure*}

Figure \ref{fig 3:methodology} shows the pipeline of the proposed model architecture. In our proposed model, there are three main components, namely FPN, neuro-synaptic dimensionality compressor, and cognitive sequence inference engine, the connections representing the flow of data through the model. We can see that the EEG signals are processed under the band-pass filter in the model. Consider that the raw data from the EEG sensors is represented as $X_{raw} \in \mathbb{R}^{n \times ch \times t }$, where $ch$ represents the number of channels, $t$ represents the number of time samples and $n$ is the number of tokens (that is, word-aligned EEG segments) and can vary between documents and subjects. It is subjected to a bandpass filter to select the frequency components and remove unwanted data. The representation of the preprocessed data is as follows:
\begin{equation}
    X_{\text{pre}} = \text{BPF}(X_{\text{raw}}, f_{\text{low}}, f_{\text{high}})
\end{equation}
$X_{\text{pre}} \in \mathbb{R}^{n \times ch \times t}$ is the filtered EEG signal, and $f_{\text{low}}$ and $f_{\text{high}}$  are the low and high cutoff frequencies of the bandpass filter. In the implementation of this framework, the values of $f_{low}$ and $f_{high}$ are 0.5 and 30 Hz. This 3-dimensional EEG data goes into the feature pyramid for further processing.
\subsubsection{Autoencoder-driven Feature Pyramid Networks}
The preprocessed data \( X_{\text{pre}} \) are then passed through an autoencoder driven FPN. To make the data compatible with dense layers in the autoencoder, the 3D matrix is flattened into a 2D matrix:
\begin{equation}
X_{\text{flat}} = \text{Flatten}(X_{\text{pre}}), \quad X_{\text{flat}} \in \mathbb{R}^{n \times d} ,\quad d = t \cdot ch
\end{equation}
Here, n is the number of samples and d is the number of features per sample. In our next step, we have implemented a model architecture that comprises three encoder layers and three decoder layers. It seems similar to traditional FPN networks, but traditional FPN networks comprise CNN layers to extract multihierarchical features from the data, whereas we have introduced a novel model that replaces CNN with an encoder-decoder layer. This replacement is introduced because autoencoder-driven FPNs are better at handling EEG noise, capturing latent sentiment features, learning temporal patterns, and subject adaptation than traditional FPNs. 
$X_{flat}$ is processed by the initial encoder layer, which reduces the original data dimensions and transmits the output to the subsequent encoder layer, while also retaining the intermediate output for skip connections with the corresponding decoder layer. This procedure is repeated for all three layers of the encoding segment. Mathematically, it can be written as:
\begin{equation}
    \mathbf{h}_1 = f_1(X_{flat}) = \text{ReLU}(\mathbf{W}_1 X_{flat} + \mathbf{b}_1), \quad \mathbf{h}_1 \in \mathbb{R}^{n \times128} \\
\end{equation}
\begin{equation}
\mathbf{h}_2 = f_2(\mathbf{h}_1) = \text{ReLU}(\mathbf{W}_2 \mathbf{h}_1 + \mathbf{b}_2), \quad \mathbf{h}_2 \in \mathbb{R}^{n \times64} \\
\end{equation}
\begin{equation} 
Z = f_3(\mathbf{h}_2) = \text{ReLU}(\mathbf{W}_3 \mathbf{h}_2 + \mathbf{b}_3), \quad Z \in \mathbb{R}^{n \times32} 
\end{equation}
The term $\mathbf{h}_1 \in \mathbb{R}^{n \times128}$ and $\mathbf{h}_2 \in \mathbb{R}^{n \times64}$ denotes the intermediate encoded representations obtained from the first and second encoder layers. The compressed latent representation, also known as the bottleneck feature, is denoted by $Z \in \mathbb{R}^{n \times32}$. The matrices $\mathbf{W}_1$,$\mathbf{W}_2$, and $\mathbf{W}_3$  are the learnable weight parameters associated with the encoder layers, while $\mathbf{b}_1$,$\mathbf{b}_2$, and $\mathbf{b}_3$ represent the corresponding bias vectors. The function $\text{ReLU}(\cdot)$ stands for the Rectified Linear Unit activation function, which is defined as $\text{ReLU}(x) = \max(0, x)$, and is applied element-wise to introduce nonlinearity into the network.\\
The decoder reconstructs the original EEG input from the latent representation $Z\in \mathbb{R}^{n \times32}$ using three fully connected sequential layers. At each stage, the dimensionality is increased, progressively approaching the original input size. To enhance information flow and preserve hierarchical features, skip connections from earlier encoder layers are integrated into the decoding process.\\
The decoding process begins with a dense layer:
\begin{equation}
    \mathbf{h}_4 = f_4(Z) = \text{ReLU}(\mathbf{W}_4 Z + \mathbf{b}_4), \quad \mathbf{h}_4 \in \mathbb{R}^{n \times64}
\end{equation}
A skip connection is then applied by adding the first encoder activation:
\begin{equation}
    \tilde{\mathbf{h}}_4 = \mathbf{h}_4 + \mathbf{h}_2
\end{equation}
This enriched representation is passed to the next decoder layer:
\begin{equation}
    \mathbf{h}_5 = f_5(\tilde{\mathbf{h}}_4) = \text{ReLU}(\mathbf{W}_5 \tilde{\mathbf{h}}_4 + \mathbf{b}_5), \quad \mathbf{h}_5 \in \mathbb{R}^{n \times128}
\end{equation}
A second skip connection is added using the encoder's second intermediate output:
\begin{equation}
    \tilde{\mathbf{h}}_5 = \mathbf{h}_5 + \mathbf{h}_1
\end{equation}
The final decoding layer then maps this representation back to the original input dimension \( \mathbb{R}^{n \times d} \):
\begin{equation}
    \mathbf{h}_6 = f_6(\tilde{\mathbf{h}}_5) = \text{ReLU}(\mathbf{W}_6 \tilde{\mathbf{h}}_5 + \mathbf{b}_6), \quad \mathbf{h}_6 \in \mathbb{R}^{n \times d}
\end{equation}
Finally, a sigmoid activation function produces the reconstructed output:
\begin{equation}
    \hat{X} = \sigma(\mathbf{h}_6)
\end{equation}
This architecture enables the decoder to integrate both abstract compressed features and detailed encoder-level features through skip connections. These residual pathways help capture fine-grained temporal-spatial dependencies inherent in EEG signals, improving the model reconstruction fidelity and downstream task performance.
\subsubsection{Neuro-Synaptic Dimensionality Reduction Unit (NSDRU)}
To effectively compress high-dimensional EEG data without loss of information, NSDRU has been implemented. The NSDRU extracts the most important EEG features while reducing spatial dimensionality, thereby improving computational efficiency and reducing the risk of overfitting.
The NSDRU processes the reconstructed EEG feature map \( \hat{X} \) through a sequence of convolutional layers, interleaved with non-linear activations and max-pooling operations. Each convolutional layer applies a set of learned filters to capture local temporal and spatial dependencies across channels.
\begin{equation}
    \mathcal{X} = \text{NSDRU}(\hat{X})
\end{equation}
where \( \mathcal{X} \) is a compressed representation that retains the most important features of $\hat{X}$. By leveraging its hierarchical convolutional structure, the NSDRU effectively captures complex spatio-temporal patterns in EEG data while compressing the feature space to half its size along each dimension.
 \subsubsection{Cognitive Sequence Inference Engine}
Following the stages of feature extraction and dimensionality compression, a recurrent neural architecture is used to capture the temporal dependencies inherent in the data. Specifically, a GRU-based network, referred to as the Cognitive Sequence Inference Engine, is utilized to model the sequential nature of the input. To enhance the representational capacity and capture long-range temporal correlations, a deep architecture is constructed comprising six stacked GRU layers. 
Unit operations are defined by:  
\begin{equation}
\left.
\begin{aligned}
z_t^{(i)} &= \sigma\left(W_z^{(i)} x_t + U_z^{(i)} h_{t-1}^{(i)} + b_z^{(i)}\right) \\
r_t^{(i)} &= \sigma\left(W_r^{(i)} x_t + U_r^{(i)} h_{t-1}^{(i)} + b_r^{(i)}\right) \\
\tilde{h}_t^{(i)} &= \tanh\left(W_h^{(i)} x_t + U_h^{(i)} \left(r_t^{(i)} \odot h_{t-1}^{(i)}\right) + b_h^{(i)}\right) \\
h_t^{(i)} &= \left(1 - z_t^{(i)}\right) \odot h_{t-1}^{(i)} + z_t^{(i)} \odot \tilde{h}_t^{(i)}
\end{aligned}
\right\}
\end{equation}
where, \( h_t^{(i)} \) denotes the hidden state of the \( i^\text{th} \) GRU unit at the time step \( t \). Each GRU unit independently models temporal dependencies within the input sequence while maintaining its own internal state. The update gate \( z_t^{(i)} \) governs the interpolation between the previous hidden state \( h_{t-1}^{(i)} \) and the candidate activation \( \tilde{h}_t^{(i)} \), effectively controlling how much of the past information is retained. The reset gate \( r_t^{(i)} \), on the other hand, modulates the contribution of the previous hidden state when computing the candidate activation, allowing the model to selectively forget irrelevant past information.
\begin{equation}
    H_t = \frac{1}{k} \sum_{i=1}^{k} h_t^{(i)}
\end{equation}
Here, \( h_t^{(i)} \) denotes the hidden state output by the \( i^\text{th} \) GRU branch at time \( t \). The architecture consists of six parallel GRU branches (that is, \( k = 6 \)), each independently capturing temporal dependencies from the input sequence. The hidden states of all branches are aggregated by computing their average, resulting in a combined hidden representation \( H_t \). The aggregated vector \( H_t \) serves as the unified hidden representation at time step \( t \), synthesizing the diverse temporal dynamics learned in all GRU branches. This averaging strategy assumes that each GRU contributes equally to the final representation, promoting a balanced ensemble of features. By consolidating information from multiple temporal perspectives, \( H_t \) captures a richer and more robust understanding of the input sequence, potentially improving performance in downstream tasks such as sequence classification.\\
A key feature of this parallel GRU architecture is the consistent dimensionality maintained across all six GRU units. This uniform structure facilitates seamless integration and aggregation of temporal representations across layers. This configuration enables the network to learn diverse, yet complementary, temporal dynamics from multiple processing pathways. As a result, the architecture produces a progressively enriched and temporally sensitive representation of the input EEG signals, improving the network's capacity to perform nuanced cognitive inference tasks.
\subsubsection{Prediction Layer}
After obtaining temporal dependencies with the Cognitive Sequence Inference Engine, the final hidden state $H_t$ is fed into a fully connected layer, followed by a softmax activation function to predict the sentiment label. Hidden states are transformed into a logit vector through a fully connected layer by the following process:
\begin{equation}
    O = W_{fc} H_t + b_{fc}
\end{equation}
where \( O \in \mathbb{R}^2 \) represents the output logits for the positive and negative sentiments, \( W_{fc} \in \mathbb{R}^{2 \times h} \) is the weight matrix of the fully connected layer, with \( h \) being the dimensionality of the hidden state obtained from the GRU, and \( b_{fc} \in \mathbb{R}^2 \) is the corresponding bias vector.
To obtain the class probabilities, a softmax function is applied:
\begin{equation}
    \hat{y} = \text{softmax}(O) = \frac{e^{o_i}}{\sum_{j=1}^{2} e^{o_j}}
\end{equation}
\( \hat{y} \in \mathbb{R}^2 \) is the vector of class probabilities and \( o_i \) is logit for class \( i \). Finally, the predicted class is determined as
\begin{equation}
    \hat{c} = \arg\max_i \hat{y}_i
\end{equation}
\( \hat{c} \) represents the predicted class label (positive or negative).
\section{RESULTS \& DISCUSSION}
 The proposed FPN-based model achieved an accuracy of 76.42.\% on the ZUCO 2.0 dataset, significantly outperforming the established baseline LSTM of 69.5\% with strong statistical support (\(p < 0.001\)). This 6.88 percentage-point improvement corresponds to a 22.3\% relative error reduction and demonstrates a large practical impact (Cohen's \(d = 2.1\)). With precision of 79.78\%, recall of 95.41\%, and an F1 score of 86.33\%, the model shows balanced and robust performance in sentiment classification. Table I presents subject-wise metrics on ZUCO 2.0, while Table II compares the proposed model performance with existing approaches on ZUCO 2.0, DEAP \cite{b22}, SEED \cite{b23}. Figure 4 illustrates the results of the subject-wise evaluation, highlighting the model's effectiveness with large-scale data and affirming its potential for real-world applications in cognitive sentiment analysis.\\
\begin{table}[]
\centering
\caption{Comparison of the performance metrics of the proposed framework on ZUCO 2.0}
\label{table:combined_metrics}
\resizebox{\columnwidth}{!}{%
\begin{tabular}{|c|c|c|c|c|}
\hline
\textbf{Subject   ID}       & \textbf{Accuracy} & \textbf{Precision} & \textbf{Recall} & \textbf{F1 Score} \\ \hline
YAC           & 0.778 & 0.813    & 0.942    & 0.861    \\ \hline
YAG           & 0.779 & 0.805  & 0.939 & 0.861 \\ \hline
YDG           & 0.783 & 0.799 & 0.939 & 0.870 \\ \hline
YDR           & 0.766 & 0.809 & 0.926 & 0.855 \\ \hline
YFR           & 0.764 & 0.809 & 0.937 & 0.864 \\ \hline
YFS           & 0.772 & 0.808  & 0.939 & 0.863 \\ \hline
YHS           & 0.776 & 0.814    & 0.939 & 0.87  \\ \hline
YIS           & 0.77  & 0.816 & 0.941  & 0.874 \\ \hline
YLS           & 0.831 & 0.831 & 0.987        & 0.907 \\ \hline
YMD           & 0.831 & 0.831 & 0.991        & 0.907 \\ \hline
YMS           & 0.696  & 0.696  & 0.989        & 0.821 \\ \hline
YRH           & 0.764 & 0.80 & 0.946  & 0.866 \\ \hline
YRK           & 0.770 & 0.808 & 0.939 & 0.872 \\ \hline
YSD           & 0.762 & 0.810 & 0.941 & 0.874 \\ \hline
YTL           & 0.772 & 0.817 & 0.934 & 0.880 \\ \hline
YAK           & 0.778 & 0.818 & 0.945 & 0.860 \\ \hline
YRP           & 0.675 & 0.675 & 0.983       & 0.806 \\ \hline
YSL           & 0.675 & 0.675 & 0.993        & 0.806 \\ \hline
\textbf{Mean} & 0.764 & 0.791  & 0.953   & 0.862   \\ \hline
\textbf{Standard Deviation} & 0.042           & 0.050           & 0.023         & 0.027          \\ \hline
\end{tabular}%
}
\end{table}
\begin{table*}[]
\centering
\small
\caption{Performance comparison over ZUCO 2.0, DEAP and SEED datasets}
\label{tab:performance_comparison}
\begin{tabular}{|c|ccc|ccc|ccc|}
\hline
\multirow{2}{*}{\textbf{Model}} &
  \multicolumn{3}{c|}{\textbf{ZUCO 2.0}} &
  \multicolumn{3}{c|}{\textbf{DEAP}} &
  \multicolumn{3}{c|}{\textbf{SEED}} \\ \cline{2-10} 
 &
  \multicolumn{1}{c|}{Accuracy} &
  \multicolumn{1}{c|}{Precision} &
  \multicolumn{1}{c|}{Recall} &
  \multicolumn{1}{c|}{Accuracy} &
  \multicolumn{1}{c|}{Precision} &
  \multicolumn{1}{c|}{Recall} &
  \multicolumn{1}{c|}{Accuracy} &
  \multicolumn{1}{c|}{Precision} &
  \multicolumn{1}{c|}{Recall} \\ \hline
Bidirectional GRU \cite{b26} &
  \multicolumn{1}{c|}{0.701} &
  \multicolumn{1}{c|}{0.710} &
  \multicolumn{1}{c|}{0.851} &
  \multicolumn{1}{c|}{0.953} &
  \multicolumn{1}{c|}{0.932} &
  \multicolumn{1}{c|}{0.929} &
  \multicolumn{1}{c|}{0.971} &
  \multicolumn{1}{c|}{0.964} &
  \multicolumn{1}{c|}{0.981} \\ \hline
Merged LSTM \cite{b27} &
  \multicolumn{1}{c|}{0.695} &
  \multicolumn{1}{c|}{0.654} &
  \multicolumn{1}{c|}{0.889} &
  \multicolumn{1}{c|}{0.582} &
  \multicolumn{1}{c|}{0.507} &
  \multicolumn{1}{c|}{0.535} &
  \multicolumn{1}{c|}{0.952} &
  \multicolumn{1}{c|}{0.958} &
  \multicolumn{1}{c|}{0.965} \\ \hline
Multimodal CNN \cite{b28} &
  \multicolumn{1}{c|}{0.708} &
  \multicolumn{1}{c|}{0.593} &
  \multicolumn{1}{c|}{0.820} &
  \multicolumn{1}{c|}{0.926} &
  \multicolumn{1}{c|}{0.985} &
  \multicolumn{1}{c|}{0.986} &
  \multicolumn{1}{c|}{0.969} &
  \multicolumn{1}{c|}{0.981} &
  \multicolumn{1}{c|}{0.985} \\ \hline
Proposed model &
  \multicolumn{1}{c|}{0.764} &
  \multicolumn{1}{c|}{0.797} &
  \multicolumn{1}{c|}{0.954} &
  \multicolumn{1}{c|}{0.967} &
  \multicolumn{1}{c|}{0.958} &
  \multicolumn{1}{c|}{0.926} &
  \multicolumn{1}{c|}{0.991} &
  \multicolumn{1}{c|}{0.989} &
  \multicolumn{1}{c|}{0.993} \\ \hline
\end{tabular}
\end{table*}
\begin{figure}[!ht] 
    \centering
    \includegraphics[width=0.5\textwidth]{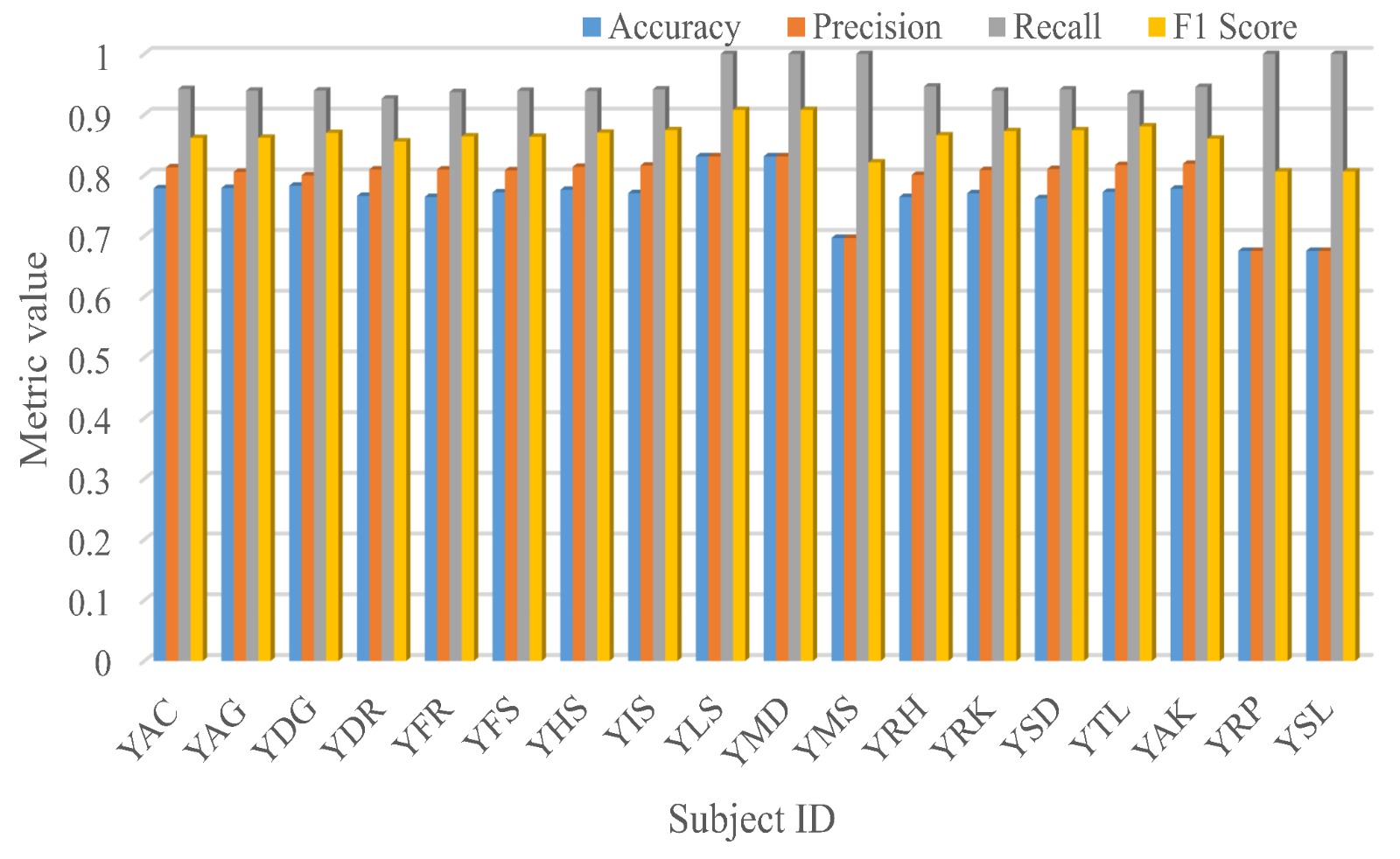} 
    \caption{Subject-wise metrics comparison on ZUCO 2.0}
    \label{fig 4:example_image}
\end{figure}
\begin{figure*}[!ht] 
    \centering
    \includegraphics[width=\textwidth]{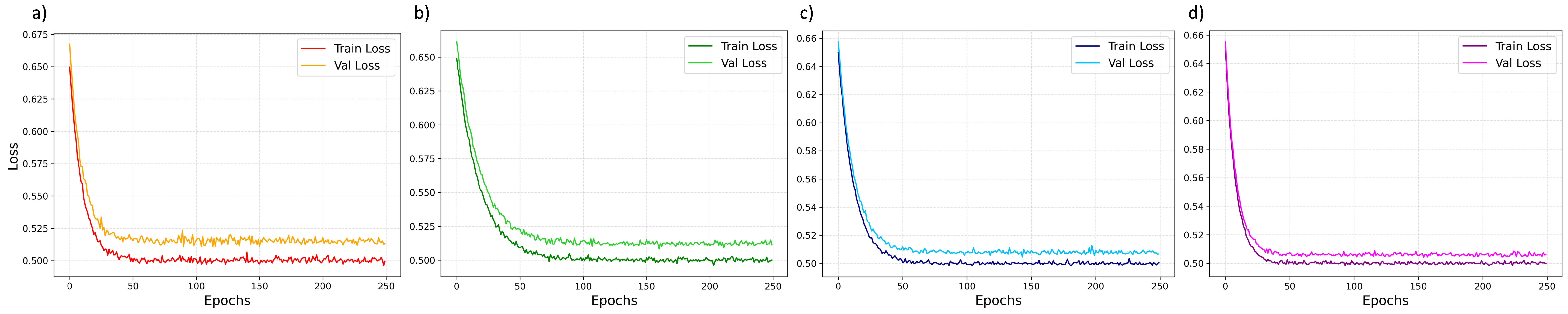} 
    \caption{Training and validation loss comparison across a) CNN b) LSTM c) GRU d) The proposed model on the ZUCO 2.0 dataset.}
    \label{fig 5:example_image}
\end{figure*}
\begin{figure}[!ht] 
    \centering
    \includegraphics[width=\columnwidth]{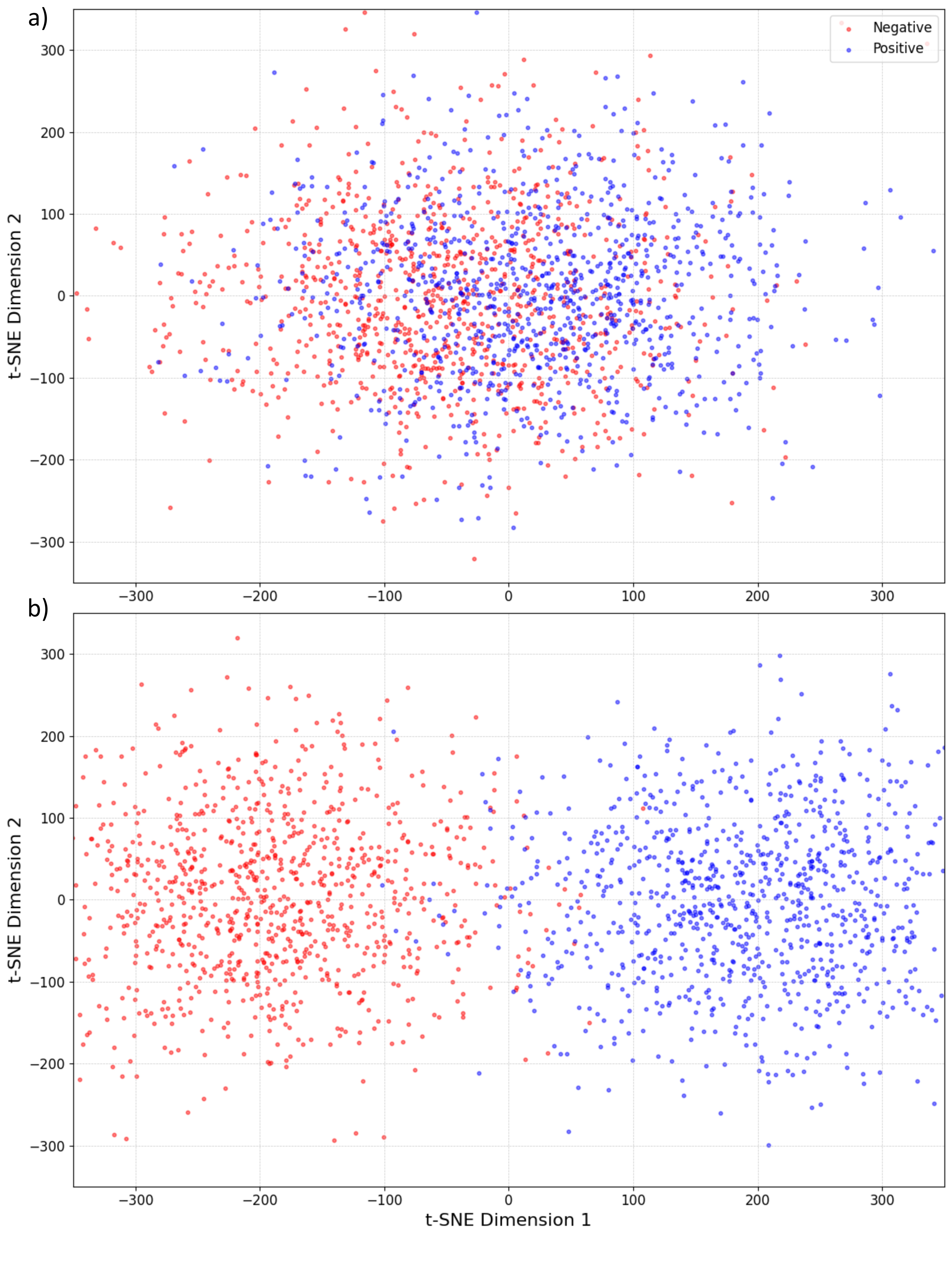} 
    \caption{t-SNE visualization of feature embeddings before (a) and after (b) applying the proposed model on ZUCO 2.0 dataset.}    
    \label{fig 7:example_image}
\end{figure}
 Figure \ref{fig 5:example_image} shows the comparative graphs of training and validation loss across CNN, LSTM, GRU, and the proposed model on the ZUCO 2.0 dataset. In these plots, it can be clearly seen that there is larger gap between validation and training loss across CNN and LSTM models, while GRU performs comparatively better with smoother convergence. The proposed model demonstrates the most stable and minimal divergence between the training and validation curves, suggesting improved learning efficiency and reduced overfitting. Figure \ref{fig 7:example_image} presents the t-SNE visualizations \cite{b24} of the feature space before and after the embedding layer of the proposed model. Initially, the positive and negative sentiment classes exhibit substantial overlap, indicating poor separability in the raw features. After embedding, the data points form clearer and more distinct clusters, demonstrating the model’s ability to extract and organize sentiment-relevant information. This progression underscores the crucial role of the embedding layer in improving the representational capacity of the model and improving sentiment discrimination. Furthermore, a detailed analysis of the DEAP and SEED datasets is described in the supplementary material.\\
For a sentimental analysis application, the most widely used classification models include LSTM, GRU, and CNN \cite{b26}\cite{b27}\cite{b28}. These models generally achieve an accuracy below 90\% for valence and arousal, indicating limitations in the capture of complex emotional states. A significant advancement is achieved with Fuzzy Convolution Logic (FCL), which integrates fuzzy logic with deep learning, yielding 98.21\% and 98.08\% accuracy for valence and arousal, respectively \cite{b13} \cite{b25}. This highlights the effectiveness of hybrid approaches in handling nuanced EEG-based emotion recognition tasks.\\
In addition to accuracy, computational efficiency is essential for the practical deployment of EEG-based models.
As detailed in Table III, the proposed model demonstrates superior computational efficiency, requiring the fewest FLOPs (507M) while maintaining a compact architecture with 1.023 million trainable parameters. It also achieves the fastest inference times on both CPU (241.3 ms) and GPU (0.137 ms). This favorable balance between model compactness and computational efficiency, achieved without compromising predictive performance, establishes the proposed framework as a highly efficient and scalable solution for EEG-to-sentiment prediction.
\begin{table}[h]
\centering
\caption{Comparison of models based on parameters, FLOPs, and inference time}
\label{tab:inference_times}
\setlength{\tabcolsep}{7pt} 
\renewcommand{\arraystretch}{1.15}
\begin{tabular}{|p{48pt}|p{36pt}|p{36pt}|p{32pt}|p{32pt}|}
\hline
\textbf{Model} & \textbf{Params} & \textbf{FLOPs} & \textbf{CPU(ms)} & \textbf{GPU(ms)} \\ 
\hline
Bidirectional GRU \cite{b26}      & 0.124M & 1,023M & 487.0 & 0.277 \\ 
\hline
Merged LSTM \cite{b27}     & 0.164M & 601M  & 286.1 & 0.162 \\ 
\hline
Multimodal CNN \cite{b28}      & 2.079M & 582M  & 276.8 & 0.153 \\ 
\hline
Proposed Model &  0.102M & 507M  & 241.3 & 0.137 \\ 
\hline
\end{tabular}
\vspace{-4.5mm}
\end{table}

\section{Conclusion}
In this paper, a novel framework is presented for EEG-based sentiment analysis that uses a GRU structured with an FPN to enhance feature extraction and temporal modelling. Our approach effectively captures multiscale representations of EEG signals, allowing for a more nuanced understanding of emotional states. The results demonstrate that our model outperforms traditional architectures, including CNNs, LSTMs, and standard GRUs, achieving significant improvements in classification accuracy, precision, recall, and F1 score. This performance gain can be attributed to the FPN’s ability to learn hierarchical features and the GRU’s efficiency in processing sequential data. Furthermore, the fast processing capabilities of our model make it suitable for real-time applications in affective computing and mental health monitoring, where timely feedback
is essential. The successful integration of a Feature Pyramid Network as part of the feature extraction process highlights the importance of capturing intricate patterns within EEG data, providing a robust foundation for sentiment analysis. Future work will focus on exploring the incorporation of attention mechanisms to further enhance the interpretability and performance of the model.

\end{document}